\documentclass[preprintnumbers,showpacs,amsmath,amssymb,floatfix,prd,onecolumn,superscriptaddress,nofootinbib]{revtex4}
\usepackage{graphicx}
\usepackage{bm}
\usepackage{amsfonts}
\usepackage{bm}
\usepackage{amsthm}
\usepackage{epstopdf}

\begin{document}

\title{Generalized Quantum Spring}

\author{Chao-Jun Feng}
\email{fengcj@shnu.edu.cn} 
\affiliation{Division of Mathematical and Theoretical Physics, Shanghai Normal University,}
\affiliation{Center for Astrophysics, Shanghai Normal University,\\ 100 Guilin Road, Shanghai 200234, China}

\author{Xiang-Hua Zhai}
\email{zhaixh@shnu.edu.cn} 
\affiliation{Division of Mathematical and Theoretical Physics, Shanghai Normal University,}
\affiliation{Center for Astrophysics, Shanghai Normal University,\\ 100 Guilin Road, Shanghai 200234, China}

\author{Xin-Zhou Li}
\email{kychz@shnu.edu.cn} 
\affiliation{Center for Astrophysics, Shanghai Normal University,\\ 100 Guilin Road, Shanghai 200234, China}

\begin{abstract}
  Recently, it was found that after imposing a helix boundary condition on a scalar field, the Casimir force coming from the quantum effect is linearly proportional to  $r$, which is the ratio of  the pitch  to  the circumference of the helix. This linear behavior of the Casimir force is just like that of the force obeying the Hooke's law on a spring. In this paper, inspiring by some complex structures that lives in the cells of human body like DNA, protein, collagen etc., we generalize the helix boundary condition to a more general one, in which the helix consists of a tiny helix structure, and makes up a hierarchy of helix.  After imposing this kind of boundary condition  on a massless and a massive scalar, we calculate the Casimir energy and force by using the so-called zeta function regularization method. We  find that the Hooke's law with the generalized helix boundary condition is not exactly the same as  usual one. In this case, the force  is proportional to the cube of $r$ instead. So we regard it as a generalized Hooke's law, which is complied by a \emph{generalized quantum spring}.
\end{abstract}


 \maketitle


\section{Introduction}\label{sec:intro}

With some boundary conditions, the dynamics and spectrum of a quantum field will be changed, and then it could lead to an observable effect. One of such a kind of phenomena is called the Casimir effect \cite{Casimir}\cite{Plunien:1986ca}. In the classical electrodynamics, the force acting between two planes doesn't exist. However, due to the quantum effect, there exists the zero-point fluctuation of the vacuum and then its spectral density changes with time. As a result, once there are two infinitely large perfectly  parallel conducting planes placed in the vacuum, there will be an attractive force between these two planes.

The Casimir effect has been studied a lot due to the development of much more precise measurements \cite{Decca:2007yb} and technological advancements since the last decade. The study on the Casimir force indeed gives some enlightenment to the nanotechnology such as the  actuation of microelectromechanical and nanoelectromechanical systems (MEMS and NEMS)\cite{MEMS}. However the same force could be generated by the stiction of devices, so there still need further works to enhance the strength and also sign of the Casimir force. It should be noticed that a repulsive force would provide an anti-"stiction" effect.

From the  theoretical aspect, the boundary conditions, material properties, temperature, and geometry that makes the Casimir force different have been studied a lot. And there are also some new methods that have been developed for computing the Casimir force between a finite number of some compact objects \cite{Emig:2007cf},  see also \cite{Li}\cite{Li11} for the case that inside a rectangular box or cavity. When the helix boundary condition was imposed on a scalar field in a flat spacetime, and when the pitch of the helix is smaller than its circumference, the Casimir force  is linearly proportional to the ratio $r$ of the pitch to the circumference, which is just like the Hooke's law that governs the force on a spring, so the authors call it \textit{quantum spring} \cite{Feng:2010qj} \cite{Zhai:2010mr} or \textit{quantum anti-spring} \cite{Zhai:2011zza} corresponding to the periodic-like and the anti-periodic-like boundary condition, see also \cite{Zhai:2011pt}. For a recent review on the Casimir effect, see \cite{Brevik:2012ht},  also see  \cite{Brevik:2001gd} for a more extensive review of the Casimir theory.

In this paper, we generalize the helix boundary condition or the quantum spring structure to a more general case, in which the helix  consists of  a tiny helix structure. In other words, it is a hierarchy of helix structure, and we call it the \emph{generalized quantum spring}, see Fig.\ref{fig:show} . More detail descriptions will be presented in the next section. In fact, there are many things living in the cells of human body, like DNA, protein and collagen having this kind of structure but even more complex. Thus, it is really interesting to find the effect of this kind of boundary condition presenting in the $(d+1)$-dimensional space-time manifold for a quantum field. Here, we impose it on a scalar field to calculate the corresponding Casimir energie and force. The method we used is called the zeta function regularization method \cite{Elizalde}, which is a very useful and elegant technique to calculate the Casimir force. Rigorous extension of the Epstein $\zeta$-function regularization and its proof have been studied in \cite{Elizalde}. The vacuum polarization of on a string was firstly discussed in \cite{Helliwell:1986hs}. Also, the generalized $\zeta$-function has been gotten many applications. For instances, see \cite{Li:1990bz}\cite{Li:1990bz11} for studying the piecewise string, see \cite{Teo:2010hr} for discussing the  noncommutative spacetime, even for the monopoles \cite{BezerradeMello:1999ge}, the p-branes \cite{Shi:1991qc} or the pistons \cite{Zhai,Zhai11,Zhai12,Zhai13,Zhai14,Zhai15,Zhai16}. Casimir effect for a fractional boundary condition has been also considered, for example, the finite temperature Casimir effect for a scalar field with fractional Neumann conditions \cite{Eab:2007zz}, while the repulsive force from fractional boundary conditions has been studied \cite{Lim:2009nk}.

This paper is organized as follows. The calculations of the Casimir energy and force under the generalized helix boundary condition for a massless and a massive scalar field in different spacetime dimensions will be presented in Sec.~\ref{sec:massless}, and Sec.~\ref{sec:massive}. A more general case is discussed in Sec.~\ref{sec:gen}, and in the last section, we will give some conclusions.

\section{Casimir force for a massless scalar field with the generalized helix boundary condition}\label{sec:massless}

The Casimir effect arises not only in the presence of some material boundaries, such as two neutral, parallel conducting plates, but also in spaces with nontrivial topology. For instance, the topology of a circle $S^1$ could cause a periodicity condition on a scalar field in one direction, e.g.  $\phi(t,x,z)=\phi(t,x+C,z)$ in $2$-dimensional space with  $(x,z)\in \mathcal{R}^2 $and $C$ the circumference of $S^1$. In other words, the boundary condition  $\phi(t,x,z)=\phi(t,x+C,z)$ can be drawn on the topology of a cylinder, while the boundary conditions $\phi(t,x,z)=\phi(t,x+a,z)$ and $\phi(t,x,z)=\phi(t,x,z+b)$ can be drawn on the topology of the torus with $a,b$ the circumference of the torus in $x$ and $z$ directions. In Ref.\cite{Zhai:2011pt}, the authors have discussed this kind of quotient topology in details.

\begin{figure}[h]
\begin{center}
\includegraphics[width=0.5\textwidth]{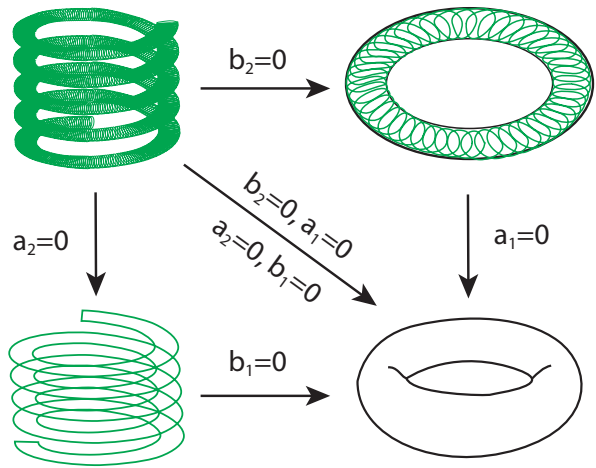}
\caption{\label{fig:show}Illustration of \emph{generalized quantum spring} and its special cases.  }
\end{center}
\end{figure}

In the following, we will  consider the generalized helix boundary condition by using the concept of quotient topology and  calculate the Casimir energy and  force for a massless scalar field  in varies spacetime dimensions. For a massive scalar field, the calculation will be presented in the next  section. As we mentioned in the above section, this kind of boundary condition is a simplification of real structures like DNA, protein, collagen etc. living in the cells of human body. To illustrate this structure, we show it in Fig.\ref{fig:show}, from which one can see that the helix consists of tiny helix structures, which makes up  a hierarchy of helix.

\subsection{A massless scalar field in the flat $2+1$ dimension}\label{sec:21dim}
In the $2+1$ dimensional spacetime with coordinates $(x,z)\in \mathcal{R}^2$, we have the following boundary condition to mimic the structure
\begin{eqnarray} \label{boudary condition}
	 \phi(t, x+a_{1}, z) &=& \phi(t,x,z+b_{1})\,,  \label{boundary condition1} \\
	 \phi(t, x+a_{2}, z) &=& \phi(t,x,z-b_{2})\,, \label{boundary condition2}
\end{eqnarray}
Where Eq.(\ref{boundary condition1}) is the helix boundary  condition with  the circumference $a_1$ and  the pitch $b_1$ of the helix. If another period condition $\phi(t, x+a_{2}, z)= \phi(t,x,z)$ or $\phi(t, x, z)= \phi(t,x,z-b_2)$ is imposed on this helix, it makes a circle in the $x$ or $z$ direction with the circumference  $a_2$ or $b_2$. Therefore, with the condition Eq.(\ref{boundary condition2}),  the helix tries to make a circle in $x$ with the circumference  $a_2$, but this time the helix will meet itself with a distance $b_2$ in the $z$ direction. In other words, the helix makes another helix with the circumference  $a_2$ and the pitch $b_1$. So in the special case ($a_2=0$ or $b_2 = 0$), Eqs.(\ref{boundary condition1}) and (\ref{boundary condition2}) reduce to the usual helix boundary condition considered in Ref. \cite{Feng:2010qj, Zhai:2010mr, Zhai:2011zza, Zhai:2011pt}, see also \cite{Feng:2012zm}.  In a even more special case ($a_2=b_1=0$ or $a_1 = b_2 = 0$), it will reduce to the torus boundary condition.

Note that we can exchange the label of coordinate $(x,z)$ since $a_1,b_1,a_2,b2$ are just some parameters,  and the result will be not changed. So we assume that $x$ is the direction along the pitch of the biggest helix structure, while $z$ is the direction along the pitch of the tiny one in the following. One can image that $x$ is the vertical direction in Fig.\ref{fig:show}, while $z$ is the azimuth coordinate.  

Under the boundary condition (\ref{boundary condition1}) (\ref{boundary condition2}), the
modes of the field are then given by
\begin{equation}\label{modes}
    \phi_{n}(t, x, z)= \mathcal{N} e^{-i\omega_nt+ik_x x+ik_z z  }\,,
\end{equation}
where $\mathcal{N}$ is a normalization factor and here, $k_x$ and $k_z$ satisfy
\begin{equation}\label{Amatrix}
	k_{x} = \frac{2\pi \big(b_{1}n_{2}+ b_{2}n_{1}\big)}{a_{1}b_{2}+a_{2}b_{1}}  \,, \qquad
	k_{z} = \frac{2\pi \big(a_{1}n_{2} - a_{2}n_{1}\big)}{a_{1}b_{2}+a_{2}b_{1}}  \,, \qquad
	n_{1}, n_{2}=0, \pm1, \pm2, \cdots \,,
\end{equation}
and also
\begin{equation}
	w_{n_{1}n_{2}}^{2} = k_{x}^{2} + k_{z}^{2} \,.
\end{equation}
Therefore, we get the energy density as
\begin{equation}\label{energy}
	E(a_{1},b_{1},a_{2},b_{2}) = \frac{1}{2} \sum^{\infty}_{n_{1},n_{2}=-\infty} \frac{2\pi}{(a_{1}b_{2}+a_{2}b_{1})^{2}}
	\sqrt{n_{1}^{2}(a_{2}^{2}+b_{2}^{2}) +n_{2}^{2}(a_{1}^{2} + b_{1}^{2}) +2n_{1}n_{2}(b_{1}b_{2}-a_{1}a_{2})} \,,
\end{equation}
see App.~\ref{def ed} for the definition of energy density here. For simplicity, here we will consider the case of $a_{1}a_{2}=b_{1}b_{2}$, and a more general case without this requirement will be discussed in the last section.  Now, the cross term in Eq.~(\ref{energy}) vanishes and we obtain
\begin{equation}
	E(a_{1},b_{1},a_{2}) = \frac{\pi b_{1}}{a_{2}^{2}(a_{1}^{2}+b_{1}^{2})^{3/2}}  \sum^{\infty}_{n_{1},n_{2}=-\infty}
	\sqrt{ n_{1}^{2}  a_{2}^{2} + n_{2}^{2} b_{1}^{2} } \,.
\end{equation}
 To calculate the above summation, we define $\mathcal{E}(s)$ as
\begin{eqnarray}
	\mathcal{E}(a_{1},b_{1},a_{2}, s) &=& \frac{\pi b_{1}}{a_{2}^{2}(a_{1}^{2}+b_{1}^{2})^{3/2}}
	\sum^{\infty}_{n_{1},n_{2}=-\infty}
	\bigg[n_{1}^{2}a_{2}^{2} +n_{2}^{2}b_{1}^{2}\bigg]^{-s/2} \\
	&=& \frac{\pi b_{1}}{a_{2}^{2}(a_{1}^{2}+b_{1}^{2})^{3/2}} \mathcal{Z}_{2}\left(a_{2}, b_{1}, s\right) \,,
\end{eqnarray}
where $\mathcal{Z}_{p}(z_{1}, \cdots z_{p};s)$ is the Epstein $\zeta$ function defined as
\begin{equation}
	\mathcal{Z}_{p}(z_{1}, \cdots z_{p};s) = \sum^{\infty }_{n_{1}, \cdots, n_{p}=-\infty} \!\!\!{}^{'}\,\,\bigg[ (n_{1}z_{1})^{2} + \cdots + (n_{p}z_{p})^{2}\bigg]^{-s/2}\,,
\end{equation}
where the prime means that the term $n_{1} = n_{2} = \cdots = n_{p} =0$ has to be excluded. Applying the reflection formulae
\begin{equation}\label{reflec}
	(z_{1}\cdots z_{p}) \Gamma\left(\frac{s}{2}\right) \pi ^{-s/2} \mathcal{Z}_{p}(z_{1}\,,\cdots \,, z_{p};s)
	= \Gamma\left(\frac{p-s}{2}\right) \pi^{(s-p)/2}  \mathcal{Z}_{p}\left(\frac{1}{z_{1}} \,, \cdots \,,  \frac{1}{z_{p}}; p-s\right) \,,
\end{equation}
and taking $s=-1$, we get the results
\begin{equation}
	E_{R}(a_{1},b_{1},a_{2}) = \mathcal{E}(a_{1},b_{1},a_{2}, -1) = -\frac{1}{4\pi a_{2}^{3}(a_{1}^{2}+b_{1}^{2})^{3/2}}  \mathcal{Z}_{2}\left(\frac{1}{a_{2}}\,,  \frac{1}{b_{1}}; 3\right) \,.
\end{equation}
By using
\begin{equation}
	Z_{2}(z_{1},z_{2};3) = \frac{2\pi^{2}}{3z_{1}^{2}z_{2}} + \frac{16\pi}{z_{1}z_{2}^{2}} \sum_{m,n=1}^{\infty} \frac{n}{m} K_{1}\left(2\pi m n \frac{z_{1}}{z_{2}}\right) + \frac{2\zeta(3)}{z_{2}^{3}} \,,
\end{equation}
we get
\begin{equation}\label{ced}
	E_{R}(a_{1},b_{1},a_{2}) =  -\frac{1}{2\pi (a_{1}^{2}+b_{1}^{2})^{3/2}} \, \mathcal{G}_{2}\left(\frac{b_{1}}{a_{2}}\right) \,,
\end{equation}
where $K_{n}(z)$ is the so-called modified Bessel function and here we have defined the function
\begin{equation}
	\mathcal{G}_{2}(x) \equiv \frac{\pi^{2}}{3} x
	+ \zeta(3)x^{3}	
	+8\pi x^{2}\sum_{m,n=1}^{\infty} \frac{n}{m} K_{1}\left(2\pi m n x \right) \,.
\end{equation}
Here, it should be noticed that, the condition $a_{1}a_{2} = b_{1}b_{2}$ is under the symmetry of $(a_{1}\leftrightarrow b_{1}, a_{2}\leftrightarrow b_{2})$,  and the Casimir energy (\ref{ced}) also respects this symmetry. And the  Casimir force along the $x$ direction is given by the following
\begin{equation}\label{force21}
	F_{a_{1}}(a_{1},b_{1},a_{2})  = - \frac{\partial E_{R}}{\partial a_{1}} =  -\frac{3}{2\pi a_1^4 (1+r_1^2)^{5/2}} \, \mathcal{G}_{2}\left(r_2\right) \,,
\end{equation}
where $r_1 = b_1/a_1$ and $r_2 = b_1/a_2$. Note that this force is along the pitch direction of the biggest helix structure.  

In  Ref.\cite{Feng:2010qj, Zhai:2010mr}, the corresponding force with the usual helix boundary condition is given by
\begin{equation}
  F_{a_{1}}(a_{1},b_{1},a_{2}) \approx -\frac{3\zeta(3) r_1}{2\pi a_{1}^{4} }  \,,
\end{equation}
in the limit of $r_1\ll 1$, which is almost linearly depending on $r_1$. So it is just like the force on a spring complying with the Hooke's law, but in this case, the force originates from the quantum effect, namely, the Casimir effect \cite{Feng:2010qj, Zhai:2010mr}. For comparing with our results, we take the limit of $r_1\ll1$ and $r_2\gg 1$, then the force (\ref{force21}) becomes
\begin{equation}\label{equ:forcelimit2}
  F_{a_{1}}(a_{1},b_{1},a_{2}) \approx -\frac{3\zeta(3) r_2^3}{2\pi a_{1}^{4} }  = -\frac{3\zeta(3)}{2\pi a_{1}^{4} } \left(\frac{a_1}{a_2}\right)^3 r_1^3 \,,
\end{equation}
which is  proportional to  $r_1^3$ with a prefactor $\left(a_1/a_2\right)^3$. So, one may regard it as a generalized Hooke's law complied by a \emph{generalized quantum spring}.

\subsection{A massless scalar field in the flat $3+1$ dimension}\label{sec:31dim}
As in the $2 + 1$ dimension case, the vacuum energy in $3 + 1$ dimension is given by
\begin{equation}\label{energy31}
	E(a_{1},b_{1},a_{2}) = \frac{b_{1}}{2 a_{2} (a_{1}^{2} + b_{1}^{2})} \int_{-\infty}^{\infty} \frac{dk}{2\pi} \sum^{\infty}_{n_{1},n_{2}=-\infty}
		\sqrt{k^{2} +  \frac{(2\pi n_{1})^{2} } {(a_{1}^{2} + b_{1}^{2})}
	     + \frac{(2\pi n_{2})^{2} b_{1}^{2}}{(a_{1}^{2} + b_{1}^{2} ) a_{2}^{2}} } \,,
\end{equation}
where we have also consider the simple case $a_{1}a_{2}=b_{1}b_{2}$. We also define
\begin{eqnarray}
\nonumber
	\mathcal{E}(a_{1},b_{1},a_{2}, s) &=& \frac{2\pi b_{1}}{(a_{1}^{2} +b_{1}^{2})^{2} a_{2}^{3}} \sum^{\infty}_{n_{1},n_{2}=-\infty}
		 \bigg[  n_{1}^{2}  a_{2}^{2}
	     + n_{2}^{2} b_{1}^{2}  \bigg]^{(1-s)/2}  \int_{0}^{\infty} dk (k^{2}+1)^{-s/2} \\
	\nonumber
	&=& \frac{\pi^{3/2} b_{1}}{(a_{1}^{2} +b_{1}^{2})^{2} a_{2}^{3}} \frac{\Gamma(\frac{s-1}{2})}{\Gamma(\frac{s}{2})}\mathcal{Z}_{2}\left(a_{2}, b_{1}, s-1\right) \\
	&=&\frac{\pi^{-1/2+s}}{(a_{1}^{2} +b_{1}^{2})^{2} a_{2}^{4}} \frac{\Gamma(\frac{3-s}{2})}{\Gamma(\frac{s}{2})} \mathcal{Z}_{2}\left( \frac{1}{a_{2}}, \frac{1}{b_{1}}, 3-s\right) \,,
\end{eqnarray}
where we have used the reflection formulae (\ref{reflec}). Taking $s=-1$, we get
\begin{equation}
	E_{R}(a_{1},b_{1},a_{2}) = \mathcal{E}(a_{1},b_{1},a_{2}, -1)
	=-\frac{1}{2\pi^{2}(a_{1}^{2} +b_{1}^{2})^{2} a_{2}^{4}} \mathcal{Z}_{2}\left( \frac{1}{a_{2}}, \frac{1}{b_{1}}, 4\right) \,.
\end{equation}
By using
\begin{equation}
	Z_{2}(z_{1},z_{2};4) = \frac{\pi\zeta(3)}{z_{1}^{3}z_{2}}
	+ \frac{8\pi^{2}}{z_{1}^{3/2}z_{2}^{5/2}} \sum_{m,n=1}^{\infty} \left(\frac{n}{m} \right)^{3/2}K_{3/2}\left(2\pi m n \frac{z_{1}}{z_{2}}\right)
	 + \frac{\pi^{4}}{45z_{2}^{4}} \,,
\end{equation}
we get
\begin{equation}\label{ced}
	E_{R}(a_{1},b_{1},a_{2}) =  -\frac{1}{\pi ^{2}(a_{1}^{2}+b_{1}^{2})^{2}}\, \mathcal{G}_{3}\left(\frac{b_{1}}{a_{2}}\right) \,,
\end{equation}
where
\begin{equation}
	\mathcal{G}_{3} (x) = \frac{\pi\zeta(3)}{2}x + \frac{\pi^{4}}{90}x^{4} + 4\pi^{2}x^{5/2}  \sum_{m,n=1}^{\infty} \left(\frac{n}{m} \right)^{3/2}K_{3/2}\left(2\pi m n x\right)\,.
\end{equation}
Therefore, the  Casimir force along the $x$ direction is given by the following
\begin{equation}\label{force31}
	F_{a_{1}} (a_{1},b_{1},a_{2})  =  -\frac{4a_{1}}{\pi^{2} (a_{1}^{2}+b_{1}^{2})^{3}} \, \mathcal{G}_{3}\left(\frac{b_{1}}{a_{2}}\right) \,.
\end{equation}
In limit of $r_1\ll1$ and $r_2\gg 1$,  Eq.~(\ref{force31}) becomes
\begin{equation}\label{equ:forcelimit2}
  F_{a_{1}}(a_{1},b_{1},a_{2}) \approx -\frac{2\pi^2 r_2^4}{45a_{1}^{5} }  = -\frac{2\pi^2 }{45a_{1}^{5} } \left(\frac{a_1}{a_2}\right)^4 r_1^4 \,,
\end{equation}
which is  proportional to  $r_1^4$ with a prefactor $\left(a_1/a_2\right)^4$. 

\subsection{A massless scalar field in the $D+1$ dimension}\label{sec:d1dim}
It is straightforward to obtain the Casimir energy for the massless scalar field in the flat $D+1$ dimensional spacetime. The energy is given by
\begin{equation}\label{energyd1}
	E(a_{1},b_{1},a_{2}) =\frac{b_{1}}{2 a_{2} (a_{1}^{2} + b_{1}^{2})} \int_{-\infty}^{\infty} \frac{d^{D-2}k}{(2\pi)^{D-2}} \sum^{\infty}_{n_{1},n_{2}=-\infty}
		\sqrt{k_{T}^{2} +  \frac{(2\pi n_{1})^{2} } {(a_{1}^{2} + b_{1}^{2})}
	     + \frac{(2\pi n_{2})^{2} b_{1}^{2}}{(a_{1}^{2} + b_{1}^{2} ) a_{2}^{2}} } \,,
\end{equation}
with $a_{1}a_{2}=b_{1}b_{2}$. Defining
\begin{equation}
	\mathcal{E}(a_{1},b_{1},a_{2}, s) =\frac{\pi b_{1}}{a_{2}^{D}(a_{1}^{2} + b_{1}^{2})^{(D+1)/2}}\sum^{\infty}_{n_{1},n_{2}=-\infty}
		 \bigg[  n_{1}^{2}  a_{2}^{2}
	     + n_{2}^{2} b_{1}^{2}  \bigg]^{(D-2-s)/2}  \int_{-\infty}^{\infty} d^{D-2}k (k^{2}+1)^{-s/2} \,,
\end{equation}
and using the mathematical identity
\begin{equation}\label{ide1}
	\int_{-\infty}^{\infty} f(x) d^{d} x = \frac{2\pi^{\frac{d}{2}}}{\Gamma(\frac{d}{2})} \int_{0}^{\infty} r^{d-1}f(r)dr \,,
\end{equation}
as well as the relation
\begin{equation}\label{ide2}
	\int_{0}^{\infty} t^{r}(1+t)^{s} dt = B(1+r, -s-r-1) \,,
\end{equation}
we get
\begin{eqnarray}
\nonumber
		\mathcal{E}(a_{1},b_{1},a_{2}, s) &=& \frac{\pi^{D/2}b_{1}}{a_{2}^{D}(a_{1}^{2} + b_{1}^{2})^{(D+1)/2}} \frac{\Gamma(-\frac{D-2-s}{2})}{\Gamma(\frac{s}{2})} \mathcal{Z}_{2}\left(a_{2}, b_{1}, s+2-D\right) \\
		&=&\frac{\pi^{s+1-D/2}}{a_{2}^{D+1}(a_{1}^{2} + b_{1}^{2})^{(D+1)/2}} \frac{\Gamma(\frac{D-s}{2})}{\Gamma(\frac{s}{2})}  \mathcal{Z}_{2}\left( \frac{1}{a_{2}}, \frac{1}{b_{1}}, D-s\right) \,,
\end{eqnarray}
where we have used the reflection formulae (\ref{reflec}). Taking $s=-1$, we get
\begin{equation}\label{renergyd1}
	E_{R}(a_{1},b_{1},a_{2}) = \mathcal{E}(a_{1},b_{1},a_{2}, -1)
	=-\frac{\Gamma(\frac{D+1}{2}) }{2\pi^{(D+1)/2}a_{2}^{D+1}(a_{1}^{2} + b_{1}^{2})^{(D+1)/2}} \mathcal{Z}_{2}\left( \frac{1}{a_{2}}, \frac{1}{b_{1}}, D+1\right) \,.
\end{equation}
To make the above result more illuminating,  it is convenient to define the auxiliary function
\begin{equation}\label{defS}
	S(m, a; s) = \pi^{-s/2}\Gamma\left(\frac{s}{2}\right)\sum_{n=-\infty}^{\infty} \bigg[\left(\frac{m}{\pi} \right)^{2} + \left(\frac{n}{a}\right)^{2}\bigg]^{-s/2} \,, \quad (Re(s)>1) \,.
\end{equation}
Its analytic continuation to the complex $s$-plane with simple poles at is given by
\begin{equation}
	S(m, a; s) = \frac{am^{1-s}}{\pi^{(1-s)/2}} \bigg[\Gamma\left(\frac{s-1}{2}\right)+ 4\sum_{n=1}^{\infty} \frac{K_{(1-s)/2}(2nma)}{(nma)^{(1-s)/2}}\bigg] \,.
\end{equation}
Therefore, we can reexpress the Epstein $\zeta$ function as
\begin{eqnarray}
\nonumber
	\mathcal{Z}_{2}(z_{1},z_{2}; s) &=& \sum_{j,k=-\infty}^{\infty}  \!\!\!{}^{'}\,\, (j^{2}z_{1}^{2} + k^{2}z_{2}^{2})^{-s/2}\\
\nonumber
	&=& \sum_{j=-\infty}^{\infty} \!\!\!{}^{'}\,\,  \sum_{k=-\infty}^{\infty} (j^{2}z_{1}^{2} + k^{2}z_{2}^{2})^{-s/2}
		+  \sum_{j=-\infty}^{\infty} \!\!\!{}^{'}\,\, (k^{2}z_{2}^{2})^{-s/2} \\
\nonumber
	&=& \frac{2\pi^{s/2}}{\Gamma\left(\frac{s}{2}\right)} \sum_{j=1}^{\infty} S(\pi j z_{1}, 1/z_{2}; s ) + \frac{2\zeta(s)}{z_{2}^{s}} \\
\nonumber
	&=& \frac{2}{z_{1}^{s-1}z_{2}}
	 \bigg[ \frac{\pi^{1/2}\zeta(s-1)\Gamma\left(\frac{s-1}{2}\right)}{\Gamma\left(\frac{s}{2}\right)}
	 + \frac{4\pi^{s/2}}{\Gamma\left(\frac{s}{2}\right)} \left(\frac{z_{1}}{z_{2}}\right)^{(s-1)/2}\sum_{j,k=1}^{\infty}  \left(\frac{j}{k}\right)^{(1-s)/2} K_{(1-s)/2}\left(2 \pi j k  \frac{z_{1}}{z_{2}}\right)
		 + \zeta(s) \left(\frac{z_{1}}{z_{2}}\right)^{s-1} \bigg] \,.
\end{eqnarray}
 Inserting this result into Eq.~(\ref{renergyd1}) yields
 \begin{equation}
	E_{R}(a_{1},b_{1},a_{2}) = -\frac{\Gamma(\frac{D+1}{2}) }{\pi^{(D+1)/2}(a_{1}^{2} + b_{1}^{2})^{(D+1)/2}} \mathcal{G}_{D}\left(\frac{b_{1}}{a_{2}}\right) \,,
\end{equation}
where
\begin{equation}
 	\mathcal{G}_{D}(x) = \frac{\pi^{1/2}\zeta(D)\Gamma\left(\frac{D}{2}\right)}{\Gamma\left(\frac{D+1}{2}\right)} x
	 + \frac{4\pi^{(D+1)/2}}{\Gamma\left(\frac{D+1}{2}\right)} x^{D/2+1}\sum_{j,k=1}^{\infty}  \left(\frac{k}{j}\right)^{D/2} K_{D/2}\left(2 \pi j k x\right)
		 + \zeta(D+1) x^{D+1} \,,
\end{equation}
Therefore, the  Casimir force on the $x$ direction is given by
\begin{equation}
	F_{a_{1}} (a_{1}, b_{1}, a_{2}) =  -\frac{a_{1}(D+1)\Gamma(\frac{D+1}{2}) }{\pi^{(D+1)/2}(a_{1}^{2} + b_{1}^{2})^{(D+3)/2}} \, \mathcal{G}_{D}\left(\frac{b_{1}}{a_{2}}\right) \,.
\end{equation}
In limit of $r_1\ll1$ and $r_2\gg 1$,  Eq.~(\ref{force31}) becomes
\begin{equation}\label{equ:forcelimit2}
  F_{a_{1}}(a_{1},b_{1},a_{2}) \approx  -\frac{(D+1)\Gamma(\frac{D+1}{2}) }{\pi^{(D+1)/2}a_{1}^{(D+2)}}\zeta(D+1) r_2^{D+1} = -\frac{(D+1)\Gamma(\frac{D+1}{2}) }{\pi^{(D+1)/2}a_{1}^{(D+2)}}\zeta(D+1) \left(\frac{a_1}{a_2}\right)^{(D+1)} r_1^{(D+1)} \,,
\end{equation}
which is  proportional to  $r_1^{(D+1)}$ with a prefactor $\left(a_1/a_2\right)^{(D+1)}$. 

\section{A massive scalar field in the flat $D+1$ dimension}\label{sec:massive}
It is also straightforward to obtain the Casimir energy for the massive scalar field in the flat $D+1$ dimensional spacetime. The energy is given by
\begin{equation}\label{energyd1}
	E(a_{1},b_{1},a_{2}, \mu) = \frac{b_{1}}{2 a_{2} (a_{1}^{2} + b_{1}^{2})} \int_{-\infty}^{\infty} \frac{d^{D-2}k}{(2\pi)^{D-2}} \sum^{\infty}_{n_{1},n_{2}=-\infty}
		\sqrt{k_{T}^{2} +  \frac{(2\pi n_{1})^{2} } {(a_{1}^{2} + b_{1}^{2})}
	     + \frac{(2\pi n_{2})^{2} b_{1}^{2}}{(a_{1}^{2} + b_{1}^{2} ) a_{2}^{2}}  + \mu^{2}} \,,
\end{equation}
with $a_{1}a_{2}=b_{1}b_{2}$. Defining
\begin{equation}
	\mathcal{E}(a_{1},b_{1},a_{2}, \mu, s) =\frac{\pi b_{1}}{a_{2}^{D}(a_{1}^{2} + b_{1}^{2})^{(D+1)/2}}\sum^{\infty}_{n_{1},n_{2}=-\infty}
		 \bigg[  n_{1}^{2}  a_{2}^{2}
	     + n_{2}^{2} b_{1}^{2}  + \frac{ \mu^{2} a_{2}^{2} (a_{1}^{2} + b_{1}^{2}) }{(2\pi)^{2}}\bigg]^{(D-2-s)/2}  \int_{-\infty}^{\infty} d^{D-2}k (k^{2}+1)^{-s/2} \,.
\end{equation}
and using Eqs.~(\ref{ide1}) and (\ref{ide2}), we get
\begin{equation}\label{massive en}
	\mathcal{E}(a_{1},b_{1},a_{2}, \mu, s) =\ \frac{\pi^{D/2} b_{1}}{a_{2}^{D}(a_{1}^{2} + b_{1}^{2})^{(D+1)/2}} \frac{\Gamma(-\frac{D-2-s}{2})}{\Gamma(\frac{s}{2})} \sum^{\infty}_{n_{1},n_{2}=-\infty}
		 \bigg[  n_{1}^{2}  a_{2}^{2}
	     + n_{2}^{2} b_{1}^{2}  +\tilde\mu^{2}\bigg]^{(D-2-s)/2}\,.
\end{equation}
where
\begin{equation}
	\tilde\mu =  \frac{ a_{2} \sqrt {a_{1}^{2} + b_{1}^{2}} }{2\pi} \mu \,.
\end{equation}
By using the relation
\begin{eqnarray}
	\nonumber
	& & \sum_{n1,n2 = -\infty}^{\infty} \bigg[ z_{1}^{2}n_{1}^{2} + z_{2}^{2}n_{2}^{2} + m^{2}\bigg] ^{-s}
	= \frac{\pi}{z_{1}z_{2}}\frac{\Gamma(s-1)}{\Gamma(s)} m^{2-2s} \\
	&+& \frac{2\pi^{s}}{z_{1}z_{2} \Gamma(s)}  \sum_{n1,n2 = -\infty}^{\infty}  \!\!\! {}^{'} \,\, m^{1-s} \left[ \frac{n_{1}^{2}}{z_{1}^{2}} + \frac{n_{2}^{2}}{z_{2}^{2}}\right]^{(s-1)/2} K_{1-s}\left(2\pi m \left[ \frac{n_{1}^{2}}{z_{1}^{2}} + \frac{n_{2}^{2}}{z_{2}^{2}}\right]^{1/2}\right) \,
\end{eqnarray}
we get the final expression
\begin{equation}\label{massive case}
E_{R}(a_{1},b_{1},a_{2},\mu) = \mathcal{E}(a_{1},b_{1},a_{2}, -1) = - \frac{\mu^{D+1}\Gamma(-\frac{D+1}{2})}{2^{D+1}\pi^{\frac{D+1}{2}}}- \frac{ \mathcal{M}(\tilde \mu /a_{2}, b_{1}/a_{2})
}{(a_{1}^{2} + b_{1}^{2})^{(D+1)/2}}\,,
\end{equation}
where the function $\mathcal{M}$ is defined as
\begin{equation}
\mathcal{M}(x, y)=  (xy)^{\frac{D+1}{2}}  \sum_{n1,n2 = -\infty}^{\infty}  \!\!\! {}^{'} \,\,  \left[ n_{1}^{2} y^{2} + n_{2}^{2}\right]^{-\frac{D+1}{4}} K_{\frac{D+1}{2}}\left(2\pi x y^{-1} \left(n_{1}^{2} y^{2} + n_{2}^{2} \right)^{1/2}\right)  \,.
\end{equation}
Thus, the  Casimir force along the $x$ direction is given by the following
\begin{equation}
	F_{a_{1}} (a_{1}, b_{1}, a_{2}, \mu) = - \frac{a_{1} \mathcal{M}(\tilde \mu/a_{2}, b_{1}/a_{2})
}{(a_{1}^{2} + b_{1}^{2})^{(D+3)/2}}  \bigg[ (D+1) - \frac{\mathcal{M}'(\tilde\mu/a_{2}, b_{1}/a_{2})}{\mathcal{M}(\tilde\mu/a_{2}, b_{1}/a_{2})} \frac{\tilde\mu}{a_{2}}\bigg]
\,.
\end{equation}
Here
\begin{equation}
\mathcal{M}_{x}(x,y) = \frac{\partial\mathcal{M}(x,y)}{\partial x} =-2\pi x^{\frac{D+1}{2}} y^{\frac{D-1}{2}} \sum_{n1,n2 = -\infty}^{\infty}  \!\!\! {}^{'} \,\, \left[ n_{1}^{2} y^{2} + n_{2}^{2}\right]^{-\frac{D-1}{4}} K_{\frac{D-1}{2}}\left(2\pi x y^{-1} \left(n_{1}^{2} y^{2} + n_{2}^{2} \right)^{1/2}\right) \,.
\end{equation}

\section{More general case}\label{sec:gen}
In this section, we will discuss a more general case, in which we do not impose the condition $a_{1}a_{2} = b_{1}b_{2}$. As before,  to get the energy density,  we  define a function
\begin{equation}
	\mathcal{F}(a, b ,c , s) = \sum_{m,n \in \mathcal{Z}} (m^{2} + amn + bn^{2} +c)^{-s} \,,
\end{equation}
where $m=n_{1}^{2}, n= n_{2}^{2}$, and
\begin{equation}
	a = \frac{2(a_{1}a_{2} - b_{1}b_{2})}{a_{2}^{2} + b_{2}^{2}}\,, \quad
	b = \frac{a_{1}^{2} + b_{1}^{2}}{a_{2}^{2}+b_{2}^{2}} \,, \quad
	c = \frac{\mu^{2}(a_{1}b_{2}+a_{2}b_{1})^{2}}{(2\pi)^{2}(a_{2}^{2} + b_{2}^{2})}\,.
\end{equation}
Then, the energy density is obtained by
\begin{equation}
	E_{R}(a_{1}, b_{1} , a_{2}, b_{2}, \mu ) = -\frac{\pi^{\frac{D-1}{2}} (a_{2}^{2} + b_{2}^{2})^{\frac{D-1}{2}}\Gamma\left(\frac{1-D}{2}\right) \mathcal{F}\left(a,b,c, \frac{1-D}{2} \right)} {(a_{1}b_{2} + a_{2}b_{1})^{D}} \,.
\end{equation}
By using the inhomogeneous Chowla-Selberg formula, see \cite{Zhai:2011zza, Zhai:2011pt}
\begin{eqnarray}
\nonumber
\Gamma(s)\mathcal{F}(a,b,c, s) &=& \Gamma(s)c^{-s} +  2\Gamma(s)\zeta_{EH}(s,c) + \frac{2^{2s}\sqrt{\pi}}{\Delta^{s-1/2}} \zeta_{EH}(s-1/2,4c/\Delta)\Gamma(s-1/2) \\
&+& 2^{s+5/2}\pi^{s}\sum_{n=1}^{\infty} n^{s-1/2}\cos(n\pi a)  \sum_{d|n} d^{1-2s}\left( \Delta + \frac{4q}{d^{2}} \right)^{1/4-s/2}
K_{s-\frac{1}{2}}\left(\pi  n \sqrt{ \Delta + \frac{4q }{d^{2}} }\right)\,,
\end{eqnarray}
where  $d$ is the divisors of $n$ and
\begin{equation}
\Delta = 4b-a^{2} = \frac{4(a_{1}b_{2} + a_{2}b_{1})^{2}}{(a_{2}^{2} + b_{2}^{2})^{2}} >0 \,.
\end{equation}
Here, we have defined
\begin{equation}
\zeta_{EH}(s,p) = \sum_{n=1}^{\infty} (n^{2}+p)^{-s} \,.
\end{equation}
Then, we get
\begin{equation}
2\Gamma(s)\zeta_{EH}(s,p) = -\Gamma(s) p^{-s}+ \sqrt{\pi}\Gamma(s-1/2) p^{-s+1/2} + 4\pi^{s}p^{-s/2+1/4} \sum_{n=1}^{\infty} n^{s-1/2}K_{s-1/2}(2\pi n\sqrt{p}) \,.
\end{equation}
Finally, we have
\begin{eqnarray}
\nonumber
\Gamma(s)\mathcal{F}(a,b,c, s) &=&  \frac{2\pi \Gamma(s-1)}{\sqrt{\Delta}} c^{1-s} + 4\pi^{s}c^{-s/2+1/4} \sum_{n=1}^{\infty} n^{s-1/2}K_{s-\frac{1}{2}}(2\pi n \sqrt{c}) \\
&+& \sqrt{c} \left(2\pi(c\Delta)^{-1/2}\right)^{s}  \sum_{n=1}^{\infty} n^{s-1}K_{s-1}(4\pi n \sqrt{c/\Delta})\\
&+& 2^{s+5/2}\pi^{s}\sum_{n=1}^{\infty} n^{s-1/2}\cos(n\pi a)  \sum_{d|n} d^{1-2s}\left( \Delta + \frac{4q}{d^{2}} \right)^{1/4-s/2}
K_{s-\frac{1}{2}}\left(\pi  n \sqrt{ \Delta + \frac{4q }{d^{2}} }\right)\,.
\end{eqnarray}
Noticed that the first term in the above equation contributes to the Casimir energy  as
\begin{equation}
- \frac{\mu^{D+1}\Gamma(-\frac{D+1}{2})}{2^{D+1}\pi^{\frac{D+1}{2}}} \,,
\end{equation}
which is the same as that in Eq.~(\ref{massive case}) and does not contribute to the Casimir force.

\section{Conclusion}
In conclusion, we have generalized the helix boundary condition to a more general case, in which the helix  consist of  a tiny helix structure, and they make up  a hierarchy of helix structure, namely, the \emph{generalized quantum spring}. This kind of boundary condition is inspired by the fact that
there are many things living in the cells of human body, like DNA, protein and collagen having this kind of structure but more complex. Thus, it is really interesting to find the effect of this kind of boundary condition presenting in the $d+1$-dimensional space-time manifold for a quantum field. In this paper, we impose it on a massless and a massive scalar field to calculate the corresponding Casimir energies and forces. The method we used is called the zeta function regularization method \cite{Elizalde}, which is a very useful and elegant technique to calculate the Casimir force. We find that  when the helix boundary condition was imposed on a scalar field in a flat spacetime, and if the pitch of the helix is smaller than its circumference, the Casimir force that comes from the quantum effect is just like the Hooke's law that govern the force on a spring. Furthermore, we find that the Hooke's law with the generalized helix boundary condition is not exactly the same as the usual one, which is proportion to the cube of ratio $r_1$ instead as comparing with the results from Ref. \cite{Feng:2010qj, Zhai:2010mr}. So, we regard it as a generalized Hooke's law complied by a \emph{generalized quantum spring}.

\acknowledgments
This work is supported by National Science Foundation of China grant Nos.~11105091 and~11047138, ``Chen Guang" project supported by Shanghai Municipal Education Commission and Shanghai Education Development Foundation Grant No. 12CG51, and Shanghai Natural Science Foundation, China grant No.~10ZR1422000.

\appendix
\section{Definition of energy density}\label{def ed}
As we known, with the periodic boundary condition (e.g. $x\sim x+a$) in $1+1$ dimensional spacetime, one has
\begin{equation}
	w_{n}= |k_{x}| = \bigg|\frac{2\pi n}{a} \bigg| \,, \quad n \in \mathcal{Z} \,,
\end{equation}
then, the vacuum energy  is defined as
\begin{equation}
E(a) = \frac{1}{2} \sum_{n} |w_{x}| \,,   \quad \text {or} \,  \quad E(a) = \frac{a}{2} \int \frac{dk_{x}}{2\pi}
\end{equation}
if $a$ is large enough, while the energy density is defined as $\rho(a) = E(a)/a$. So, there is a correspondence between the summation and integration as the following
\begin{equation}\label{one dim}
\sum_{n} \sim a \int \frac{dk}{2\pi} \,, \quad \text{or} \, \quad \frac{1}{a}\sum_{n} \sim \int \frac{dk}{2\pi} \,,
\end{equation}
for defining of the energy and energy density.
In the case of  multiple boundary conditions $\bold{k} = \bold{A}  2\pi \bold{n}$ with $\bold{A}$ the  transition matrix,   $1/a$ in Eq.~(\ref{one dim}) should replace with the Jacobian determinant, namely, the energy density is defined as
\begin{equation}
\rho = \frac{1}{2} \det{\bold{A}}  \int \frac{d^{d}k}{(2\pi)^{d}} \sum_{\bold{n}} w_{ \bold{n}} \,,
\end{equation}
in $D+1$ dimensional spacetime. Here, $d$ denotes the number of non-quantized directions and then $A$ is a  $(D-d)\times(D-d)$ matrix.
For example, if we have two periodic boundary condition ($x\sim x+a$, $y\sim y +b$), then
\begin{equation}
A = \left(
  \begin{array}{cc}
    1/a & 0 \\
      0             &1/b \\
  \end{array}
\right)     \,,
\end{equation}
 and the energy density is defined as
 \begin{equation}
\rho = \frac{1}{2 ab}  \int \frac{dk}{2\pi} \sum_{n,m}\sqrt{k_{z}^{2} + \left(\frac{2\pi n}{a}\right)^{2}+ \left(\frac{2\pi m}{a}\right)^{2}}\,,
\end{equation}
in $3+1$ dimensional spacetime. From Eq.~(\ref{Amatrix}),  we have
\begin{equation}
 \left(
  \begin{array}{c}
    k_{x}  \\
    k_{z}  \\
  \end{array}
\right)
= \frac{1}{a_{1}b_{2}+a_{2}b_{1}}  \left(
  \begin{array}{cc}
   b_{2}  &  b_{1}\\
   -a_{2} &  a_{1} \\
  \end{array}
\right)
=  \left(\begin{array}{c}
    2\pi n_{1}  \\
    2\pi n_{2} \\
  \end{array}
\right)         \,,
\end{equation}
and $\det \bold A = 1/(a_{1}b_{2}+a_{2}b_{1})$. So, the energy density is defined as Eq.~(\ref{energy}).

\end{document}